# PID-EVOA: Tuning PID Controller Parameter Optimzation based on the Habits of the Egyptian Vulture

Mohamed Issa
Faculty of Engineering
Zagazig University
Zagazig, Egypt
Mohmed.issa@gmail.com

Abou Ella Hassanien
Faculty of Computer and Information
Cairo University
Cairo, Egypt
Aboitcairo@gmail.com

Ahmed Abd Elbaset
Faculty of Engineering,
Zagazig University
Zagazig, Egypt
ahmed_abdelbasit94@hotmail.com

*Abstract*— Proportional - Integral - Derivative (PID) Controller is a primary component in industrial control systems nowadays. Its gain parameters have a powerful effect on its transient response criteria such as integral squared error (ISE), settling time, rise time and overshooting. Tuning gain parameters of PID to reach the optimal transient response of control systems is a hard problem. The standard method is Ziegler-Nichols (ZN) method that initially computes the values of the parameters which produce reasonable response but can be enhanced. In this paper Egyptian Vulture Optimization Algorithm (EVOA) is used to tune the parameters of PID controller by minimizing the ISE function more than that using ZN. The results show that using EVOA enhanced the performance of the controller response better than ZN method when implemented to liquid level control system. The ISE was reduced from 32 to 8.9 and the settling time from 189 to 81 second.

**Keywords : PID, Liquid level control system, Meta-Heuristics, EVOA**

## I. INTRODUCTION

PID controller is a common used controller in industrial applications since its robustness and simplicity. The use of the PID algorithm does not guarantee optimal control of the system or even its stability. It's not guaranteed to work correctly; noticeably it may be affected by delays (the calculated error doesn't come immediately, or the control action does not apply instantaneously). The response of the controller can be described according to an error, the degree to which the system overshoots a set point, and the degree of oscillation of any system. But it is broadly applicable since a PID controller relies only on the measured process variable, not on knowledge of the underlying process; it has a long history of successful use in a wide range of applications [1-6].

Optimization of the transient response of control systems include s minimizing overshooting or decreasing settling time and rise time. Also, ISE also an objective to be minimized. Optimization is the process of searching for the optimal values of system's parameters from the feasible values for maximizing or minimizing its output. Meta-heuristics algorithm is a stochastic strategy mimics its behavior from social and environmental systems used to optimize engineering problems. Meta-heuristics succeed in optimizing a lot of applications in different fields such as bioinformatics [7, 8], image processing applications [9-13] and renewable energy [14, 15]. Hence meta-heuristic can be used to choose the best combination of parameters of PID controller that achieve the most efficient transient response. In this paper Egyptian Vulture optimization algorithm[16] used to tune the parameters of PID controller and compared the result with that produced by Ziegler-Nicolas method [17].

The rest of the paper is organized as follow : Section II explains the basics of PID controller while EVOA are explained is section III. Section IV explains tuning parameters of PID controller using EVOA. Section V shows experimental results of tuning parameters of PID using EVOA for three tank liquid level control. Finally the work is concluded in section VII.

## II. PID CONTROLLER

PID controller is a feedback controller used commonly in industrial control systems to provide stable state at desired output specified by the operator. As shown in Fig.1 PID controller consist of three parts proportional, integral, and derivative to control process plant, where r(t) is the desired value (set point), and y(t) is the actual measured output.

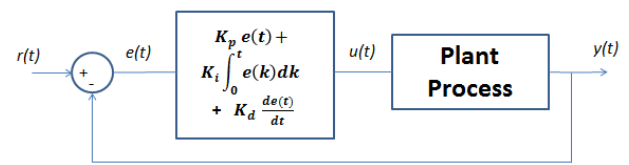

Fig.1 Block diagram of a PID controller in a feedback loop

The objective of PID controller is minimizing the error over time using the input control signal u (t) to the process plant which is computed according to eqn. (1). The process control signals u(t) such as a damper, control valve or a voltage to heating or cooling element. $K_p$, $K_i$, and $K_d$ are the coefficients of the proportional, integral and derivative terms of controller respectively.

$$u(t) = K_p\, e(t) + K_i \int_0^t e(k)dk + K_d\, \frac{de(t)}{dt} \quad (1)$$

The performance of PID controller is controlled by settling time and maximum peak. Settling time ($t_s$) is the time elapsed from the application of an ideal instantaneous step input to the time at which the output has entered and remained within a specified error band and its indication of system stability. Maximum Peak ($M_p$) is by how much a signal or function exceeds its stable target. The transient response that shows $t_s$ and $M_p$ is as shown in Fig.2.

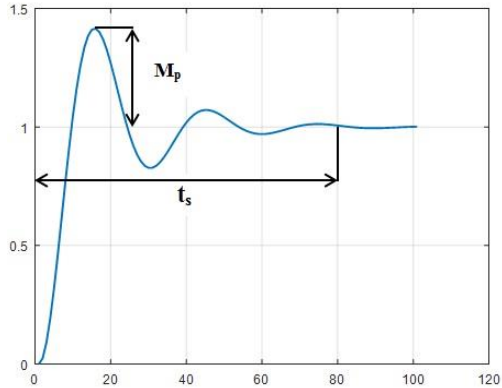

Fig.2 Transient response of PID controller.

Another important factor that determine the performance of PID controller is ISE and is determined according to eqn. (2).

$$ISE = \int_0^\infty (r(t) - h(t))^2 \quad (2),$$

where r(t) and h(t) is the desired output and actual output respectively. Hence, as ISE, $M_p$ and $t_s$ short as possible the performance is the being optimal.

The classical tuning method is formally known as the Ziegler–Nichols (ZN) method. It is performed by setting the I (integral) and D (derivative) gains to zero and the P (proportional) gain is then increased (from zero) until it reaches the ultimate gain $K_u$, at which the output of the control loop has stable and consistent oscillations Tu.

The parameters gain of various controllers are set based on $K_u$ and $T_u$ as in Table.I.

Table I
ZIEGLER-NICOLAS FORMULA FOR P, PI AND PID CONTROLLERS PARAMETERS

| Control Type | $K_p$ | $K_i$ | $K_d$ |
|---|---|---|---|
| P | 0.50 $K_u$ | | |
| PI | 0.45 $K_u$ | 0.54 $K_u$ / $T_u$ | |
| PID | 0.60 $K_u$ | 1.2 $K_u$ / $T_u$ | 3 $K_u T_u$ / 40 |

The Ziegler-Nichols tuning produce a "quarter wave decay" which gives an acceptable result for some purposes, but not optimal for all applications. Hence, more tunning for parameters of PID is required using meta-heuristics to tune the PID to be suitable for most of application.

### III. EGYPTIAN VULTURE OPTIMIZATION (EVO) ALGORITHM

EVO is a optimization technique that mimic its startegy from egyptian vulture for finding solutions of optimization problems. It can be applied in constrained and unconstrained dependent problems. There is no any paramters for EVOA need to be tuned which consider an advantage of it. It is fully randomized with no dependency on other solutions in contrast with other meta-heuristic algorithms. EVOA was used in common optimization problems such as knapsack problem [16] and travel sales man problem [16]. Algorithm 1 shows the procedure of EVOA.

**Algorithm 1 Egyptian Vulture Optimization**
1: Initialize single solution (P).
2: **Repeat**
3:   Tossing of pebbles
4:   Rolling of the solutions
5:   Change angle of tossing
6:   Evaluate fitness function and update best solution ($P_{gbest}$)
7: **Until** (T < maximum number of iterations)
8: Return the best solution ($P_{gbest}$) obtained as the global optimum

### IV. EVOA-PID CONTROLLER

The parameters of PID controllers ($K_p$, $K_i$, and $K_d$) are tuned using meta-heuristics by using the values that are derived using ZN method and the meta-heuristics try to optimize these parameters. This developed PID controller is called Heuristics PID controller, and it's rule of work as shown in Fig. 3.

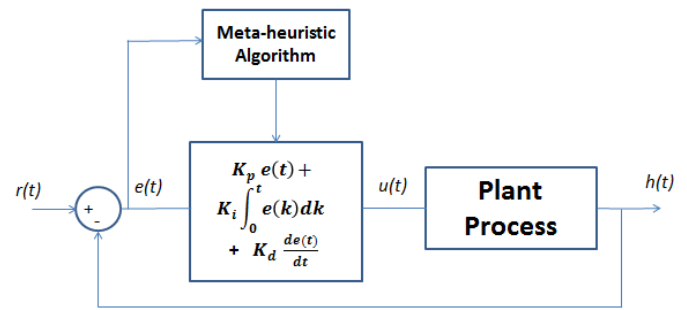

Fig.3. Block-level shows the rule of using meta-heuristic algorithm for tuning PID

#### A. Representation of solutions

$K_p$, $K_i$ and $K_d$ are continuous parameters and it was represented in this work in digital form (W-bits) and EVOA was performed this form. So decimal to discrete converting operation and the reverse are used before and after calling EVOA.

## B. PID optimization using EVOA

Fig. 4 describe the procedure of finding the optimum parameters of PID controller. The procedures initializes with the parameters values that were produced using Ziegler-Nicolas method as the best solutions and another solution with three parts is randomized within specific range of $K_p$, $K_i$ and $K_d$. EVOA then is used to generate new solutions from the previous solutions based on three steps are tossing of pebbles, rolling the solution and changes angle of tossing.

- Tossing of pebbles is performed as in Fig. 5 (a) where random position bit (P) is chosen from W-bits. From this position for L-bit (L is length of bits that represent size of pebbles) the data in these bits are changed using exclusive or logic operation with random generated L bits.
- Rolling the solution is performed as in Fig. 5 (b) to change the region of search space to increase diversity which increase probability of reaching global optimum. The rolling is simulated by rotating the solution bits left or right randomly for random times.
- Changing angle of tossing as in Fig. 5 (c) is simulated by mutated random number of bits with random positions. The benefit of this step is increasing the diversity.

The objective function was to find PID parameters gains that achieve minimum settling time and minimum ISE according to Eqn. (2).

## C. Computational Analysis

If tossing of pebbles, rolling of solution and change of angle consume time on average is $(C_{xor} + C_r + C_m) \times \frac{N}{2}$ where $C_{xor}$, $C_r$ and $C_m$ represent unit time for xor operation of 2-bits, rotation solution 1-bit and mutating 1-bit respectively. So, the time complexity is represented as the following equation :

$$O((C_{xor} + C_r + C_m) \times \frac{W}{2} \times N) \quad (3),$$

where N is the number of population of EVOA.

## V. EXPERIMENTAL RESULTS

A simple three tanks liquid level system is used as a process plant that is as shown in Fig.6. The system is modeled as the transfer function as shown in (4).

$$\frac{H(S)}{R(S)} = \frac{1}{64\,S^3 + 9.6\,S^2 + 0.48\,S + 0.008} \quad (4),$$

where H(S) and R(S) current height of liquid level and the desired height level in the frequency domain.

The criteria that were used in the EVOA-PID validation are :

- Statistical mean: is the average of solutions that are produced from executing the optimization algorithm for M times and is calculated according to (5) .

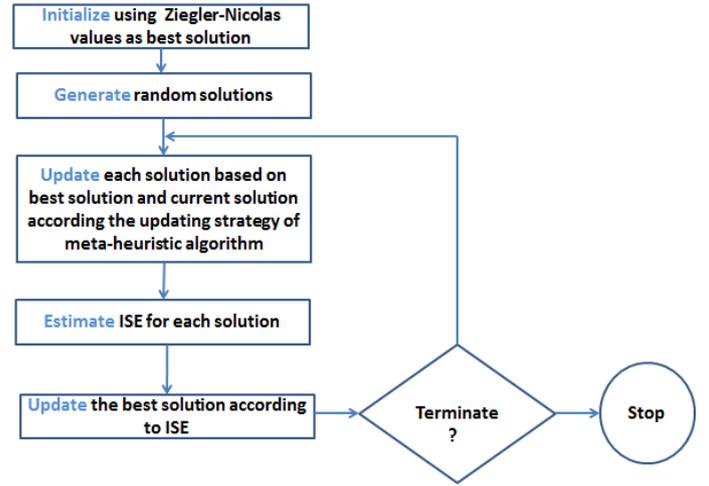

Fig. 4. Flowchart shows the steps of tuning PID parameters using meta-heuristics.

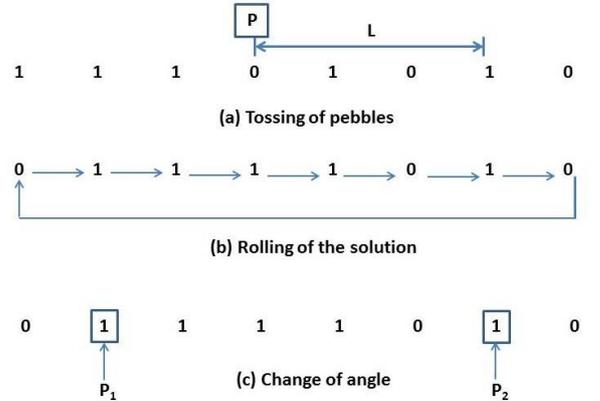

Fig. 5. EVOA operation for PID parameter optimization

$$Mean = \frac{1}{M} \sum_{i=1}^{M} S_i \quad (5)$$

where $S_i$ is the optimal solution of the run time $i$.

- Statistical standard deviation (std): is indicator for the variation of the best fitness values found for running the optimization algorithm for M run times. Also, it represents the robustness and stability and it is computed as in (6) :

$$Std = \sqrt{\frac{1}{M-1} \sum_{i=1}^{M} (S_i - Mean)^2} \quad (6)$$

Table 2 show the comparison between tuning parameters using EVOA and ZN. As shown the results of using EVOA

outperform ZN by producing minimum ISE and short settling time with small standard deviation. Fig. 7 show the transient response of liquid level system of using parameters that were tuned by ZN versus EVOA. Fig. 8 show the speed of convergence for best ISE found by EVOA versus iteration.

Table II
PID Gain Parameters Optimized Using EVOA vs ZN

| Method | $K_p$ | $K_i$ | $K_d$ | ISE | $t_s$ (Sec) | $M_p$ % |
|---|---|---|---|---|---|---|
| ZN | 0.038 | 0.001 | 0.170 | 32.7 | 189.0 | 41 |
| EOVA | 0.098 | 0.006 | 2.01 | 8.85 | 81.27 | 42 |

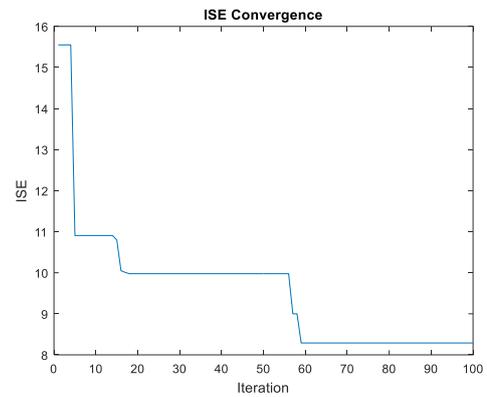

Fig. 8. Convergence speed of best ISE.

Fig. 8. Transient Response of the liquid systems by EVO vs ZN

CONCLUSION

This paper proposed Egyptian Vulture Optimization (EVO) Algorithm for tuning the parameters of PID controller by minimizing the ISE function. The experimental results show that the proposed EVO is very competitive compared with the result with that produced by Ziegler-Nichols method. PID-EVOA was tested on controlling level of liquid in three tank liquid level system. The system response was enhanced by minimizing ISE from 32 to 8.9 and settling time from 189 to 81 second which reflect power of EVOA for optimizing parameter of PID controller.

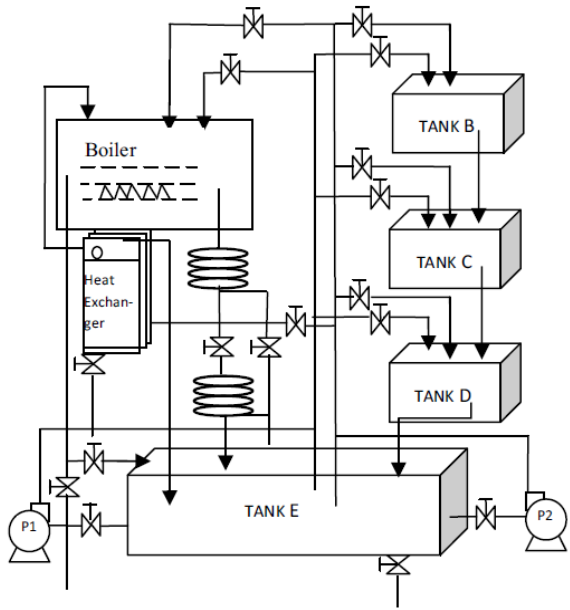

Fig. 6. Three tank liquid level system.

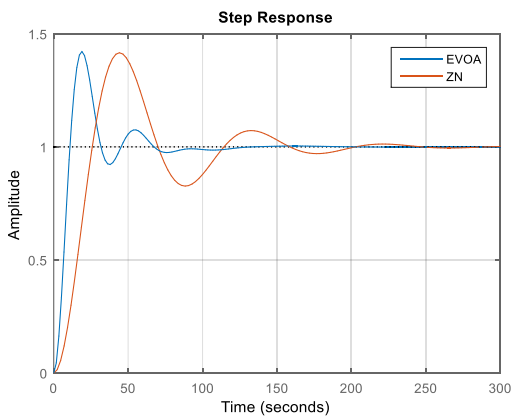

Fig. 7. Transient Response of the liquid systems by EVO vs ZN

REFERENCES

[1] Eason, G., B. Noble, and I. Sneddon, On certain integrals of Lipschitz-Hankel type involving products of Bessel functions. Philosophical Transactions of the Royal Society of London A: Mathematical, Physical and Engineering Sciences, 1955. 247(935): p. 529-551.

[2] Maxwell, J.C., A treatise on electricity and magnetism. Vol. 1. 1881: Clarendon press.

[3] Jacobs, I., Fine particles, thin films and exchange anisotropy. Magnetism, 1963: p. 271-350.

[4] Zhang, Y., et al., Data-Driven PID Controller and Its Application to Pulp Neutralization Process. IEEE Transactions on Control Systems Technology, 2017.

[5] Chen, Y., F.-Y. Xia, and B.-X. Xu. Study of PID Control Algorithm and Intelligent PID Controller. in Mechatronics and Automation Engineering: Proceedings of the International Conference on Mechatronics and Automation Engineering (ICMAE2016). 2017. World Scientific.

[6] Wang, M., et al., An artificial immune system algorithm with social learning and its application in industrial pid controller design. Mathematical Problems in Engineering, 2017. 2017.

[7] Ali, A.F. and A.-E. Hassanien, A Survey of Metaheuristics Methods for Bioinformatics Applications, in Applications of Intelligent Optimization in Biology and Medicine. 2016, Springer. p. 23-46.

[8] Issa, M. and A.E. Hassanien, Multiple Sequence Alignment Optimization Using Meta-Heuristic Techniques, in Handbook of Research on Machine Learning Innovations and Trends. 2017, IGI Global. p. 409-423.

[9] El Aziz, M.A., A.A. Ewees, and A.E. Hassanien, Whale Optimization Algorithm and Moth-Flame Optimization for multilevel thresholding image segmentation. Expert Systems with Applications, 2017. 83: p. 242-256.

[10] Elfattah, M.A., et al. Handwritten Arabic Manuscript Image Binarization Using Sine Cosine Optimization Algorithm. in International Conference on Genetic and Evolutionary Computing. 2016. Springer.

[11] Hassanien, A.E., et al. Historic handwritten manuscript binarisation using whale optimisation. in Systems, Man, and Cybernetics (SMC), 2016 IEEE International Conference on. 2016. IEEE.

[12] El Aziz, M.A., A.A. Ewees, and A.E. Hassanien, Hybrid swarms optimization based image segmentation, in Hybrid Soft Computing for Image Segmentation. 2016, Springer. p. 1-21.

[13] Said, S., et al. Moth-flame Optimization Based Segmentation for MRI Liver Images. in International Conference on Advanced Intelligent Systems and Informatics. 2017. Springer.

[14] Oliva, D., M.A. El Aziz, and A.E. Hassanien, Parameter estimation of photovoltaic cells using an improved chaotic whale optimization algorithm. Applied Energy, 2017. 200: p. 141-154.

[15] Oliva, D., et al., A Chaotic Improved Artificial Bee Colony for Parameter Estimation of Photovoltaic Cells. Energies, 2017. 10(7): p. 865.

[16] Sur, C., S. Sharma, and A. Shukla. Egyptian vulture optimization algorithm–a new nature inspired meta-heuristics for knapsack problem. in The 9th International Conference on Computing and InformationTechnology (IC2IT2013). 2013. Springer.

[18] Ziegler, J.G. and N.B. Nichols, Optimum settings for automatic controllers. trans. ASME, 1942. 64(11).